\documentclass[12pt,preprint]{aastex}

\shorttitle{Pre-transitional disk nature of the AB Aur disk}
\shortauthors{Honda et al.}

\begin{document}
\title{Pre-transitional disk nature of the AB Aur disk \altaffilmark{1}}


\author{M. Honda\altaffilmark{2}, A. K. Inoue\altaffilmark{3},
Y. K. Okamoto\altaffilmark{4}, H. Kataza\altaffilmark{5},
M. Fukagawa\altaffilmark{6}, T. Yamashita\altaffilmark{7}, 
T. Fujiyoshi\altaffilmark{8}, 
M. Tamura\altaffilmark{7},
J. Hashimoto\altaffilmark{7}, T. Miyata\altaffilmark{9},
S. Sako\altaffilmark{9}, 
I. Sakon\altaffilmark{10},
H. Fujiwara\altaffilmark{5,10}, T. Kamizuka\altaffilmark{5,10} and T. Onaka\altaffilmark{10} 
}


\altaffiltext{1}{Based on data collected at Subaru Telescope, which is operated
by the National Astronomical Observatory of Japan.}
\altaffiltext{2}{Department of Information Sciences, Kanagawa University,
2946 Tsuchiya, Hiratsuka, Kanagawa, 259-1293, Japan}
\altaffiltext{3}{College of General Education, Osaka Sangyo University,
3-1-1, Nakagaito, Daito, Osaka 574-8530, Japan}
\altaffiltext{4}{Institute of Astrophysics and Planetary Sciences,
Faculty of Science, Ibaraki University, 2-1-1 Bunkyo, Mito, Ibaraki 310-8512, Japan}
\altaffiltext{5}{Department of Infrared Astrophysics
Institute of Space and Astronautical Science (ISAS)
Japan Aerospace Exploration Agency (JAXA)
3-1-1 Yoshinodai, Chuo-ku, Sagamihara, Kanagawa 252-5210, Japan}
\altaffiltext{6}{Department of Earth and Space Science, Graduate School
of Science, Osaka University, 1-1 Machikaneyama, Toyonaka, Osaka
560-0043, Japan}
\altaffiltext{7}{National Astronomical Observatory of Japan, 2-21-1
Osawa, Mitaka, Tokyo 181-8588, Japan}
\altaffiltext{8}{Subaru Telescope, National Astronomical Observatory of
Japan, 650 North A'ohoku Place, Hilo, Hawaii 96720, U.S.A.}
\altaffiltext{9}{Institute of Astronomy, School of
Science, University of Tokyo, 2-21-1 Osawa, Mitaka, Tokyo 181-0015, Japan}
\altaffiltext{10}{Department of Astronomy, School of Science, University
of Tokyo, Bunkyo-ku, Tokyo 113-0033, Japan}


\begin{abstract}
\noindent
The disk around AB Aur was imaged and resolved at 24.6\,$\mu$m using 
the Cooled Mid-Infrared Camera and Spectrometer on the 8.2m Subaru Telescope.
The gaussian full-width at half-maximum of the source size is estimated
 to be 90 $\pm$ 6 AU, indicating that the disk extends further out 
at 24.6\,$\mu$m than at shorter wavelengths.
In order to interpret the extended 24.6\,$\mu$m image, we consider 
a disk with a reduced surface density within a boundary radius $R_c$,
which is motivated by radio observations that suggest a reduced inner region 
within about 100 AU from the star.
Introducing the surface density 
reduction factor $f_c$ for the inner disk, 
we determine that the best match with the observed radial 
intensity profile at 24.6\,$\mu$m is achieved with $R_c$=88 AU and $f_c$=0.01.
We suggest that the extended emission at 24.6\,$\mu$m is due to the enhanced emission from 
a wall-like structure at the boundary radius (the inner edge of the outer disk),
which is caused by a jump in the surface density at $R_c$.
Such reduced inner disk and geometrically thick outer disk structure can 
also explain the more point-like nature at shorter wavelengths.
We also note that this disk geometry is qualitatively similar to a pre-transitional disk,
suggesting that the AB Aur disk is in a pre-transitional disk phase.
\end{abstract}


\section{Introduction}
The evolution of protoplanetary disks has attracted attention because 
it is presumably a site of the on-going planet formation. Mostly by
analyzing spectral energy distributions (SEDs), many transitional disks
whose inner region is devoid of small dust grains have been
detected. Recently, \cite{espaillat07} introduced a new sub-class called
pre-transitional disks with definite near-infrared (NIR) excesses that 
indicate the presence of an inner disk separated from an
optically thick outer disk. They suggested that such a disk is forming a
gap and when the inner disk dissipates, it will become a transitional disk with an inner
cavity. In both cases, the presence of a wall-like
structure on the inner edge of the outer disk can explain
the excess mid-infrared (MIR) emission seen in the SED.
For intermediate-mass young stars such as Herbig Ae/Be stars, some
objects such as HD 135344B \citep{grady09} and HD 100546 \citep{benisty10}
are thought to be in the pre-transitional or transitional disk phase.

AB Aur (d=144$^{+23}_{-17}$pc; A0 Ve, \cite{vandenAncker97}) has
been regarded as a prototypical Herbig Ae/Be star and is one of the best-studied
Herbig Ae stars, with a mass of 2.4$\pm$0.2 M$_\odot$ and an age of
4$\pm$1 Myr \citep{DeWarf03}. Previous observations at various
wavelengths revealed complex circumstellar structures at various
scales. Optical and NIR coronagraphic direct images showed an extended
circumstellar nebulosity of the envelope and spiral structures in the
disk in scattered light \cite[]{grady99,fukagawa04}. Recent
coronagraphic polarimetric imaging observations revealed a dust annulus
at a radius of $\sim$100AU \cite[]{oppenheimer08,perrin09,hashimoto10} with a
possible fluctuation in the azimuthal direction.
Millimeter and sub-mm interferometric observations showed that the gas and dust 
have been depleted in the central region ( $r <$100 AU) of the disk 
\cite[]{pietu05,lin06}.

At MIR wavelengths, many direct imaging and interferometric
studies have been performed toward AB Aur.
\cite{chen03} imaged AB Aur at 11.7 and 18.7\,$\mu$m using the Keck I
telescope, and resolved it at 18.7\,$\mu$m. They found that models with
a disk surrounded by an envelope fit their data better than
flared disk models.
Applying a deconvolution technique to the 20.5\,$\mu$m data
obtained at the 3.6 m Canada-France-Hawaii-Telescope (CFHT), \cite{pantin05}
discovered a ring-like structure at 280 AU.
\cite{marinas06} presented 11.6 and 18.5\,$\mu$m images using
the Gemini North telescope; however, they could not find the ringlike structure
reported by \cite{pantin05}. They derived the intrinsic source full-width half-maximum (FWHM) to
be 17$\pm$4 AU at 11.6\,$\mu$m and 22$\pm$5 AU at 18.5\,$\mu$m, assuming
a gaussian brightness distribution. 
\cite{liu05,liu07} also resolved the disk at 10.3\,$\mu$m by nulling
interferometric observations and derived the diameter of the disk to be
24-30 AU. 

In this paper, we present results of the first imaging observations of AB Aur at 24.6\,$\mu$m
using the 8.2m Subaru telescope. At 24.6\,$\mu$m, the point-spread function
(PSF) is stable
compared to that at shorter wavelengths because of the larger Fried length, which
enables us to observe small extended structures with high reliability. 
In addition, it allows us to trace the cooler part of the disk.
By constructing a simple model, we discuss the disk structure of AB Aur
as well as implications for the nature of the AB Aur disk.

\section{Observations and Data Reduction}
We made imaging observations with the Cooled Mid-Infrared
Camera and Spectrometer \citep[COMICS;][]{kataza00,okamoto03,sako03} on the 8.2 m
Subaru Telescope on Mauna Kea, Hawaii on 2003 October 12.
AB Aur was observed using the Q24.5-OLD filter ($\lambda_{\nu_{mean}}$=24.6\,$\mu$m,
$\Delta\lambda$=1.9\,$\mu$m). The plate-scale of COMICS is 0.13 arcsec
per pixel. The chopping throw was 10 arcsec and the
position angle (PA) of the chopping direction was 30 degree. The chopping frequency was 0.45 Hz, 
and the total integration time was 2005 s.
Just before and after AB Aur observations, we observed 
$\alpha$ Tau as the PSF reference and photometric standard star. 
The total integration times were 401 s and 251 s, respectively.

For data reduction, we employed a shift-and-add method to improve the
spatial resolution. The imaging data consist of 0.983 s on-source
integration frames. 
First, the thermal background and the dark current
signals were removed through the subtraction of the chopped pair
frames. The object was sufficiently bright to be recognized even in 0.983 s
integration chop-subtracted frames, so we searched for the peak position by
fitting a Gaussian with the source profile. Then we shifted the frames (in order to
align the peak position) and summed them, but we excluded the
frames whose gaussian FWHM deviated more than 1 $\sigma$ from
the mean value. This rejection of lower quality data resulted in an
effective integration time of 1333 s for AB Aur.
The same procedure was applied to the data of $\alpha$ Tau taken before and
after AB Aur observations, and the effective integration times were
reduced to 327 s and 209 s, respectively. 
Flux calibration was performed using the $\alpha$ Tau template spectrum
provided by \cite{engelke06}. Standard aperture photometry was performed
and the resultant flux density of AB Aur was 43.0 $\pm$ 3.4 Jy.

\section{Results}
\subsection{Source size of AB Aur at 24.6\,$\mu$m}
\begin{figure}
\epsscale{.70}
\plotone{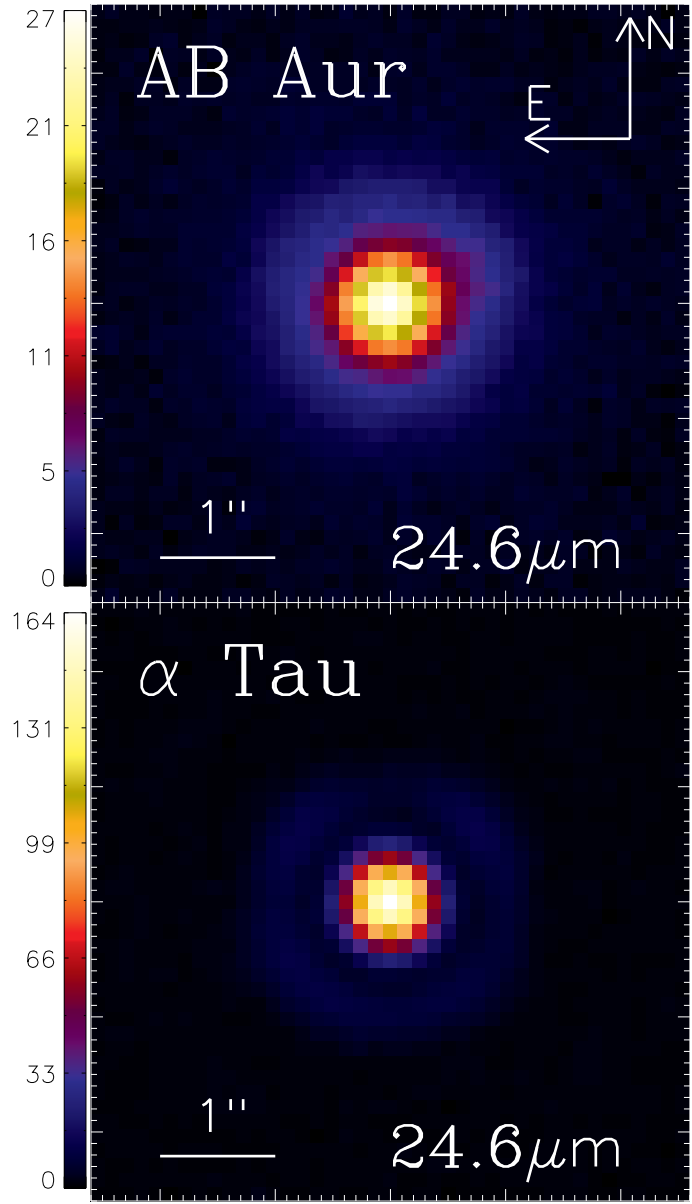}
\caption{Shift-and-added image of AB Aur (top) and $\alpha$ Tau as a PSF reference 
(bottom) at 24.6\,$\mu$m. The surface brightness is given in Jy/arcsec$^2$ and scaled from 0 to the peak
 value of each image. North is up and east is
 to the left for the AB Aur image; however, the PSF image is aligned to match the
 same instrument rotation angle as the AB Aur image for a similar optical
 configuration.
\label{fig1}}
\end{figure}

The 24.6\,$\mu$m images of AB Aur and $\alpha$ Tau are shown
in Fig.\ref{fig1}. It is clear that AB Aur is extended beyond the PSF at this
wavelength. The disk emission appears symmetric and we could not find
clear evidence of the radial profile depending on the azimuthal angle within the uncertainty.
This is probably because the disk around AB Aur is nearly face-on
\citep[inclination $\sim$ 30$^\circ$; ][]{fukagawa04}.
The FWHMs of shift-and-added images of AB Aur, as well as $\alpha$ Tau before and after
observations of AB Aur were 0.82''$\pm$0.03'', 0.647''$\pm$0.007'' and
0.644''$\pm$0.004'', respectively. This error is the standard deviation of the selected frames.
The FWHM of the PSF reference
($\alpha$ Tau) is comparable to the predicted value of the
diffraction-limited performance of the telescope at 24.6\,$\mu$m.
We estimate a rough source size of AB Aur by comparing the observed AB Aur 
image with a circular gaussian image convolved with the observed PSF image.
The best-match gaussian FWHM is 90$\pm$6 AU. 
This value is larger than that at shorter wavelengths, 
since the gaussian FWHMs of the AB Aur disk are repoted to be 17$\pm$4 AU at 11.6\,$\mu$m 
and 22$\pm$5 AU at 18.5\,$\mu$m \citep{marinas06}.

\subsection{Comparison with a simple model}
Gaussian brightness distributions are used for
the sake of simplicity and they are not a good representation of the true
intrinsic radial brightness profile of the disk. 
Thus, it would be useful to compare the observed AB Aur image with 
a disk model for the investigation of physical properties of the disk.
We can calculate the model brightness distribution of a disk if the temperature and density 
structure of the disk is given. Here we derive the disk structure 
based on a 1+1D radiative transfer model similar to that in 
\cite{dullemond02}; the stellar radiation transfer is solved 
by the so-called grazing angle approximation, but the radiative
transfer of the dust thermal radiation is solved exactly along the disk
vertical axis with a variable Eddington factor method. The dust and gas are assumed 
to be thermally and dynamically coupled, and 
the vertical density structure is determined by the hydrostatic 
equilibrium that is maintained consistent with the temperature structure. Scattering is 
not taken into account. When the disk is optically thick, the model 
predicts a flared geometry and reproduces the spectrum of protoplanetary disks. 
When the disk becomes optically thin, the grazing angle approximation does not work. 
In such cases, we treat the stellar radiation transfer by a 2D ray-tracing
method. This modification is found to work in the present calculation.
We assume dust consisting of spherical astronomical silicate grains \citep{draine84}
and the dust size distribution given by $n(a)\propto
a^{-3.5}$, where $a$ is the grain radius \citep{mathis77}. 
As \cite{perrin09} discussed that the maximum grain size is about 1\,$\mu$m, we set
the minimum and maximum grain sizes to be 0.005 $\mu$m and 2.5 $\mu$m, respectively. 
The stellar parameters used are: the
effective temperature T$_*$ = 9,500 K, the radius R$_*$ = 2.5 R$_\odot$, and the
mass M$_*$ = 2.4 M$_\odot$ \citep{vanboekel05}.
We assume a power-law surface density $\Sigma=\Sigma_1 (r/{\rm AU})^{-p}$
between $r_{\rm in}$ and $r_{\rm out}$, where p=1 \citep{hartmann98}, $\Sigma_1=70$ g cm$^{-2}$, 
$r_{\rm in}=0.5$ AU and $r_{\rm out}=1000$ AU.
The surface density is the total (gas + dust) one and we also assume a gas-to-dust mass ratio 
of 100. We chose $\Sigma_1$ and $r_{\rm out}$ so as to reproduce 
the observed far-infrared to sub-mm/mm flux densities and 
$r_{\rm in}$ is the radius where the dust temperature becomes about 1,400 K.
In addition, we introduce the boundary radius $R_c$ that separates the inner and outer
regions of the disk because the inner region (within $70-110$ AU of the central star) 
is reported to show a deficiency in the mm/sub-mm dust continuum emission \citep[]{pietu05,lin06}, 
which is attributed to a decrease in the surface density of dust grains smaller than mm-size 
in the inner region. To account for this information in our disk modeling, we consider that the
surface density is reduced by a scaling factor $f_c$ of 0.1, 0.01, and 0.001 
for $r<R_c$ compared to the power-law surface density of the outer disk. 
The explored parameter space is summarized in Table 1.
This reduction of the inner surface density makes the inner region much thinner than the 
outer disk, and as a result, a wall-like structure appears at the boundary radius. 
The upper part of the wall is directly illuminated by the central star; thus 
its temperature is enhanced like a surface super-heated layer 
\citep{chiang97}. On the other hand, the lower part of the wall 
remains cool because it is shadowed by the inner region \citep{mulders10}.
In practice, we calculate the disk structures of the inner and outer regions separately, 
and connect them at the boundary radius $R_c$.
The temperature of the illuminated upper part of the wall (i.e. inner 
edge of the outer region) is set to be the same as the surface layer 
at the wall. In addition, the same treatment is adopted at the most 
inner edge of the disk which is also directly illuminated by the central 
star. We need 2D radiative transfer calculations to be self-consistent 
\citep{mulders10}, which will be done in the future.

On the basis of the derived disk structure, we calculate the model brightness distribution for 
various wavelengths and spectrum by a ray-tracing method along the
observer's line of sight. The inclination angle is assumed to be
30$^\circ$ \citep{fukagawa04}. 
To compare the model brightness distribution with our
observation, we create a model image where the model brightness
distribution is convolved with the observed PSF image shown
in Fig.~1. Finally, we find that the best match with the observed radial profile is obtained 
with the scaling factor of the inner disk $f_c=0.01$ and the boundary radius $R_c=88$ AU.
This is comparable to the inner radius derived by \cite{pietu05} ($\sim$110 AU for dust and $\sim$70 AU for CO).
Fig.~2 shows the best-fit model brightness distribution at 25\,$\mu$m, 
the model image for Subaru/COMICS and its peak-normalized radial profile. 
Structural parameters of the best-fit model such as the radial distributions of the surface 
density and height, are shown in Fig.~3. The surface height was determined by the height where the stellar 
Planck mean opacity became unity towards the central star.
The spectrum of the best-fit model also matches with that of AB Aur from 10\,$\mu$m to $\sim1$ mm 
\citep[][and references therein]{helou88,meeus01,acke04} as shown in Fig.~4.
For wavelengths shorter than 10 $\mu$m, our model spectrum underestimates
the observed spectrum of AB Aur. This may be due to the emission 
from the inner-most gas disk \citep{tannirkulam08} or from the puffed-up inner rim \citep{natta01}, 
neither of which is considered in our model. The contribution of these inner structures
to the 25\,$\mu$m flux density is negligible; 
the spectrum at 1--8\,$\mu$m can be well fitted by an additional blackbody of temperature 1,400 K (the upper solid line)
and this blackbody component emits 1.4 Jy at 25\,$\mu$m, only 3\% of the 25\,$\mu$m Infrared 
Astronomical Satellite (IRAS) flux density of 48 Jy (see Fig.~4).

\section{Discussion}
\subsection{Comparison with other observations and validity of the wall-like structure}
A reduction of the surface density inside the boundary radius ($R_c\sim88$ AU) and the resultant wall-like
structure at $R_c$ can explain the extended emission we observed at 24.6\,$\mu$m. 
Additionally, this interpretation is consistent 
with the mm/sub-mm interferometric observations of AB Aur.
Furthermore, it can explain the properties of the AB Aur disk that appear more point-like at shorter MIR wavelengths.
In order to determine the extent of the AB Aur disk at these shorter MIR wavelengths from our best-fit model,
we calculated 11.4 and 18 \,$\mu$m model brightness distributions of
this model, and convolved
them with a gaussian model PSF comparable to the diffraction limit of an 8.2m telescope.
Then we measured FWHMs of the PSF convolved model images and 
derived gaussian FWHMs of 8 AU at 11.4\,$\mu$m and 
21 AU at 18\,$\mu$m by using the quadratic subtraction method.
The value at 18\,$\mu$m is consistent with previous observational results by \cite{marinas06} and 
that at 11.4\,$\mu$m is slightly narrower than the observations \citep{marinas06}, but the general trend is fine.
The more compact nature of the AB Aur disk at the shorter MIR wavelengths relative to 25\,$\mu$m is 
due to not only the radial 
temperature distribution but also the temperature 
of the wall-like structure (96 K). This temperature is too low for dust to radiate efficiently 
at shorter wavelengths, resulting in little change of the
appearance of the disk at shorter wavelengths.
A similar situation is also seen in another Herbig Ae/Be star HD 142527, 
which is only marginally extended at 10\,$\mu$m, but clearly extended at
24.5\,$\mu$m \citep{fujiwara06}. 

Another piece of collateral evidence for the wall-like structure at $\sim88$ AU is 
the agreement between the temperature (93K) of the blackbody 
component derived from the spectrum of AB Aur \citep{meeus01} and
the temperature (96K) of the wall at 88 AU in our best-fit disk
model. \cite{meeus01} suggested that this temperature might reflect 
the dust temperature at the onset of disk flaring.
Their suggestion is qualitatively similar to ours
in the sense that the wall in our model can be interpreted as the onset of the outer disk flaring.

Furthermore, the annulus of dust at $\sim$100 AU is observed in the NIR polarimetric 
coronagraphic images \citep{oppenheimer08,perrin09,hashimoto10}.
Inside the dust annulus of the NIR scattered light, the gap is reported at a radius 
of 0.5''$\sim$0.6'' \citep[72$\sim$86AU;][]{hashimoto10}. 
This also supports the onset of the disk flaring and the 
presence of a wall-like structure at $\sim$88 AU in the AB Aur disk.

\subsection{The AB Aur disk: a pre-transitional disk?}

Combined with the relatively young age of the central star and the
pristine environment as indicated by the global envelope
 structure, AB Aur has been thought to be a prototypical example of newly formed Herbig
 Ae/Be star \citep[e.g.][]{grady99}.
However, the inner region of the AB Aur disk possesses little gas and
 dust as revealed by mm/sub-mm interferometric observations
 \citep{pietu05,lin06}, indicating the evolved nature of the disk.
Such an apparently paradoxical situation may have led to the difficulty faced
by \cite{chen03} in explaining the AB Aur properties with a simple flared disk model.
A reduction in the inner disk surface density such as that used in our model can solve this
problem. This situation is qualitatively similar to pre-transitional disks 
introduced for some T Tauri stars \citep{espaillat07} in the sense 
that an inner disk separated from the optically thick outer disk 
is present. We propose that the AB Aur disk is also a pre-transitional disk.

Various mechanisms for the clearing of the inner region have been suggested such as 
dust coagulation \citep[e.g.][]{dullemond05}, inside-out evacuation by the 
magnetrotational instability \citep[e.g.][]{chiang07,suzuki09}, photoevaporation \citep[e.g.][]{alexander06}, 
dynamical effects of companions \citep[e.g.][]{varniere06a,varniere06b}, and so on. 
In the case of AB Aur, it is not clear which mechanism is the most plausible; 
however, the presence of planetary companions has been suggested, 
which may have caused the inner hole \citep[e.g.][]{pietu05,Jang-Condell10,hashimoto10}.
The pre-transitional disk nature of the AB Aur disk may reflect ongoing planet formation.

\acknowledgments

We are grateful to all of the staff members of the Subaru Telescope for
providing us the opportunities to perform these observations and for
their support. 
This research was partially supported by KAKENHI (Grant-in-Aid for Young
Scientists B: 21740141) by the Ministry of Education, Culture, Sports,
Science and Technology (MEXT) of Japan.

\begin{deluxetable}{cc}
\tabletypesize{\scriptsize}
\tablecaption{Range of Parameters Explored in Model Grid}
\tablewidth{0pt}
\tablehead{
\colhead{$\Sigma$ scaling factor $f_c$} & 
\colhead{Boundary radius $R_c$ [AU]}
}
\startdata
0.1   & 75, 81, 88, 95, 102, 110, 119, 128 \\
0.01  & 65, 70, 75, 81, 88, 95, 102, 110, 119 \\
0.001 & 33, 38, 44, 52, 60, 65, 70, 75, 81, 88, 95, 102 \\
\tableline
\enddata
\end{deluxetable}

\begin{figure}
\epsscale{1.0}
\plotone{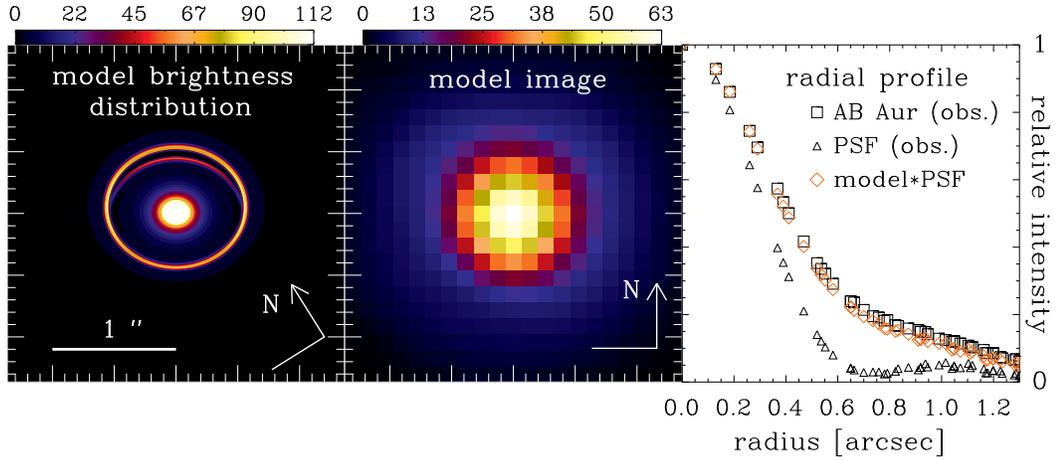}
\caption{Plots comparing the observations and the best-fit model
(boundary radius $R_c$ = 88 AU and scaling factor $f_c$ = 0.01). 
The left panel represents the model brightness distribution at 25\,$\mu$m in Jy/arcsec$^2$.
The middle panel shows the model image where the model brightness
 distribution is binned to the COMICS plate-scale (0.13''/pixel) and convolved with the observed
 PSF shown in Fig.1; additionally, the image was rotated so that the disk major-axis 
position-angle was 58$^\circ$ \citep{fukagawa04}.
The right panel shows peak-normalized azimuthally-averaged radial profile plots of the observed AB Aur image
 (squares), the PSF image (triangles), and the binned model image convolved with the PSF 
shown in the middle panel (diamonds);
horizontal axis represents radius in arcseconds.
\label{fig2}}
\end{figure}

\begin{figure}
\epsscale{0.7}
\plotone{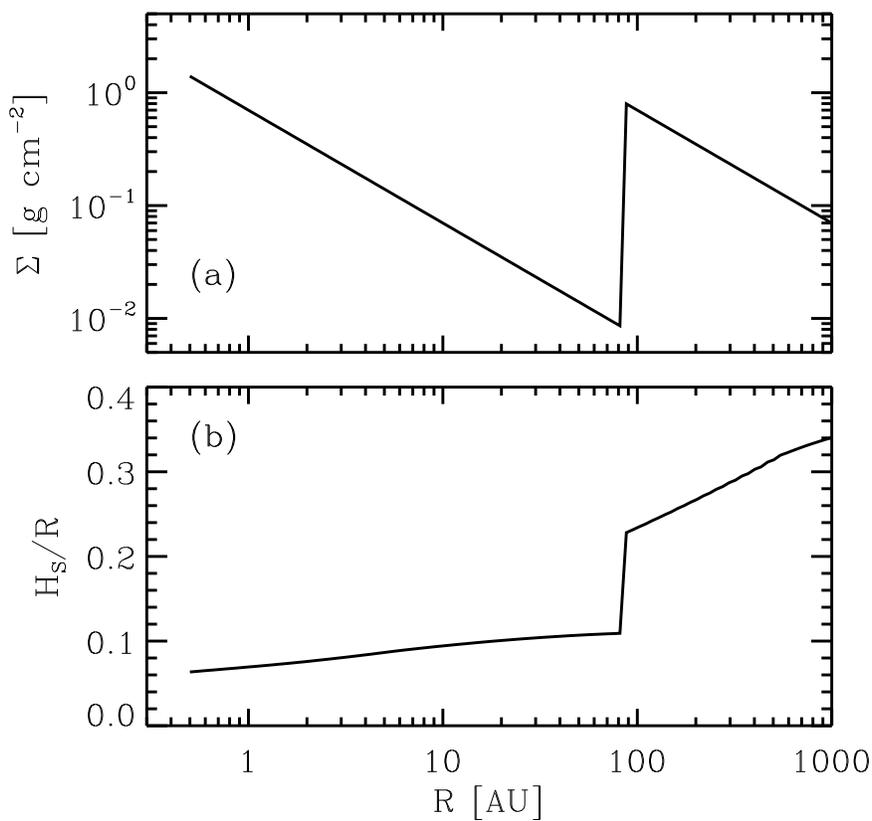}
\caption{(a) Radial distribution of the summed gas and dust surface density 
for the best-fit model (boundary radius $R_c$ = 88 AU and 
scaling factor $f_c$ = 0.01). 
(b) Radial distribution of the surface height $H_s$ normalized by R for the best-fit model.
The surface height was determined by the height where the stellar Planck mean opacity became unity 
along the line of sight to the central star. A wall-like structure is present at the boundary radius $R_c$.
\label{fig3}}
\end{figure}

\begin{figure}
\epsscale{1.0}
\plotone{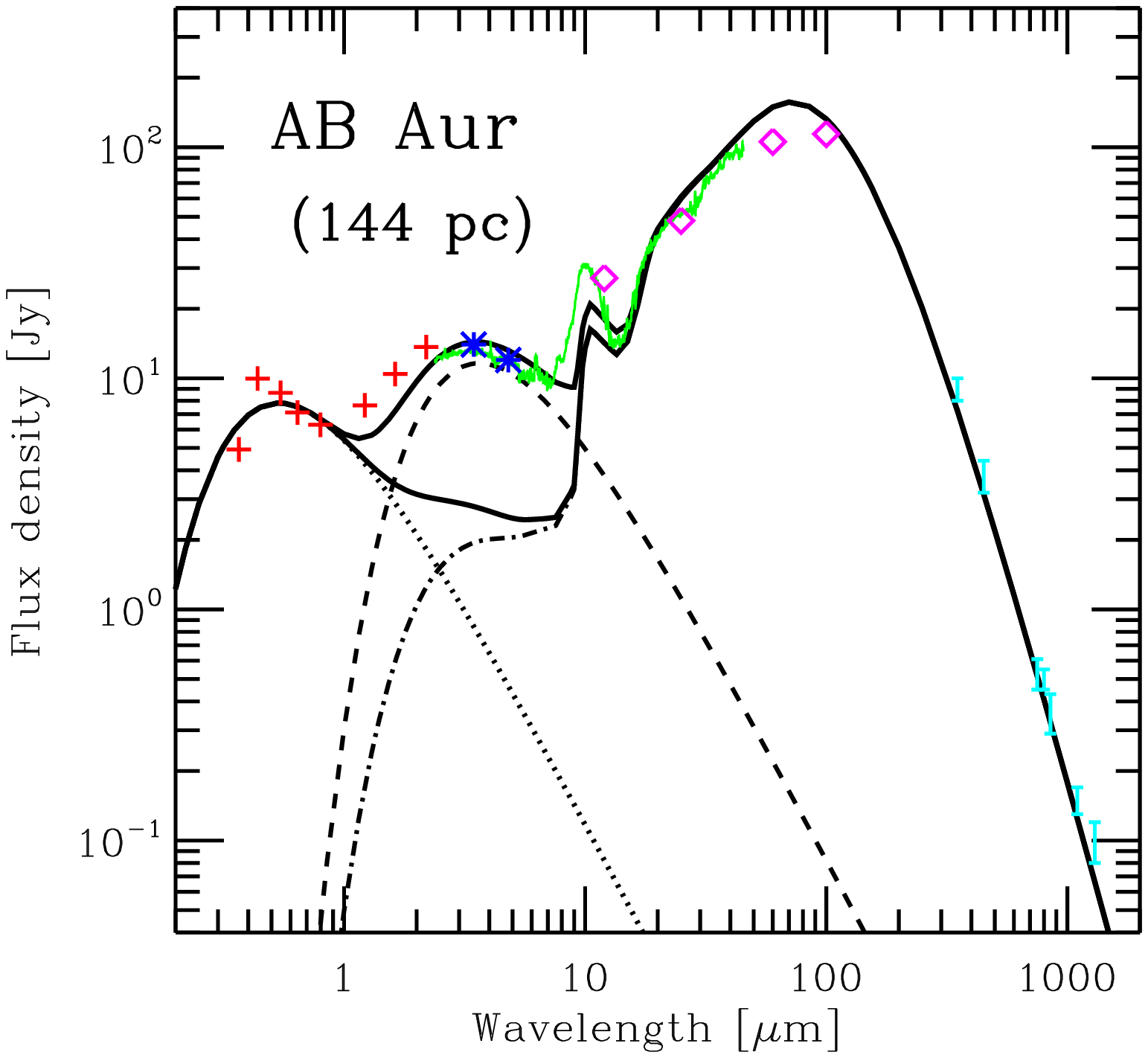}
\caption{Spectrum of AB Aur observed and that of our best-fit model. 
Observed data are from \cite{helou88,meeus01,acke04} and references therein.
Reddening correction is applied using Av=0.5 mag \citep{acke04} 
assuming the extinction law of \cite{cardelli89}.
The stellar emission is shown in dotted line. 
The dot-dashed line represents emission from the dust disk of our best-fit model.
The dashed line is for the additional 1400K blackbody needed to fit the observed spectrum, 
which is not considered in our best-fit model.
The lower solid line is a sum of the emission from the star and the dust disk, 
while the upper solid line is a sum of the star, disk and 1400K blackbody emission.
\label{fig4}}
\end{figure}

\end{document}